\documentclass[aps,prl,twocolumn,showpacs,superscriptaddress,groupedaddress,nofootinbib]{revtex4}  
\usepackage{graphicx}  
\usepackage{dcolumn}   
\usepackage{bm}        
\usepackage{amssymb}   

\begin{document}

\widetext


\title{A Generalization of Gravity
}

\author{
Chethan Krishnan}\email{chethan.krishnan@gmail.com}
\affiliation{Center for High Energy Physics, Indian Institute of Science, Bangalore, India}

\date{\today}

\begin{abstract}
I consider theories of gravity built not just from the metric and affine connection, but also other (possibly higher rank) symmetric tensor(s). The Lagrangian densities are scalars built from them, and the volume forms are related to Cayley's hyperdeterminants. The resulting diff-invariant actions give rise to geometric theories that go beyond the metric paradigm (even metric-less theories are possible), and contain Einstein gravity as a special case. Examples contain theories with generalizeations of Riemannian geometry. The 0-tensor case is related to dilaton gravity. These theories can give rise to new types of spontaneous Lorentz breaking and might be relevant for ``dark" sector cosmology.


\end{abstract}

\maketitle

\section{Introduction}

The principle of equivalence requires that the metric is flat Lorentzian and the affine connection is zero, in a freely falling frame. In metric theories of gravity, the simplest way to arrange this is to take the connection to be Riemannian (ie., that of Levi-Civita) and demand that the derivatives of the metric also vanish in the free-fall frame. That such a frame exists in Riemannian geometry is easily demonstrated: Riemann normal coordinates will do the job. 

In this paper we will construct theories where the restriction that the metric derivatives  have to vanish in the free-fall frame will be relaxed, while still requiring that the connection coefficients are zero. The basic idea is an old one going back to Riemann himself \footnote{\label{riem}I thank M. S. Narasimhan for pointing out that the higher tensor idea dates back to Riemann.}, namely that of letting the geometry depend on higher rank tensor fields instead of the metric. 

However, a dynamical metric that is relevant for geometry is crucial if one wants to have the possibility of connecting the theory to special relativity, so in this paper we will not completely sacrifice the metric. Some key features of our apporoach are: 
\begin{itemize}

\item We work with a {\em coupled} system containing the metric, higher rank symmetric tensors and the affine connection. 

\item We work in the context of a diff-invariant {\em dynamical} (ie., action-based) theory of geometry, whose construction we will describe. 
\end{itemize}

In these theories, by going to the geodesic frame, it becomes possible to bring the metric to the Minowski form as well as arrange the vanishing of the connection, while having the extra flexibility of new tensor fields in the theory. 
There will typically be a non-trivial background value for the higher rank tensor field in this frame, that spontaneously breaks Lorentz invariance. But this is a solution-dependent feature of the resulting theory, which one might (in principle) attempt to ameliorate or exploit. We will briefly discuss some salient features of these theories in the concluding section.

Various modifications and generalizations of general relativity have been reported in the literature over the last century, too many to be listed here in toto. Instead we will refer the reader to the Wikipedia page on the subject \cite{wiki} which has a pretty elaborate sample of references. See also \cite{Nojiri}.

\section{Ingredients}

One way to motivate our approach is to look for a generalization of the Levi-Civita condition
\begin{eqnarray} \label{Levi-Civita}
\nabla_a g_{bc}=0
\end{eqnarray}
in a way that incorporates higher tensors along with the metric. The primary problem in trying to find a generalized Levi-Civita condition is that the number counts for the independent components on either side do not match if one includes higher rank fields. If one assumes that the connection is torsion-free\footnote{This is an assumption we will make throughout the paper. Torsion is a tensor field on the manifold, and adding it might be interesting for various pruposes, including coupling to spinors \cite{Penrose}, but is not crucial for the conceptual line of this paper.}, both sides of (\ref{Levi-Civita}) contain the same number of independent components, $d^2 (d+1)/2$ in $d$ dimensions. This seems indeed like a precarious balance, so we might be tempted to give up any hope of generalizing the Levi-Civita condition\footnote{In principle, the connection is a choice, and can even be completely independent of the other fields in the theory, and does not necesarily have to match any number counting constraint (See \cite{Mann} for a clean discussion of this.). But we would like to find a fairly ``natural" generalization of the Levi-Civita prescription, which has some dynamical significance.}. But motivated by the fact that the relationship between the spin connection and the vielbeins does get modified in higher spin theories\footnote{Our original motivation to look for generalizations of Levi-Civita was in the context of a spacetime description (as opposed to frame bundle description) of higher spin theories. See for example section 3.1 of \cite{Campoleoni}, for some inspiration in this direction. The theories we will present in this paper only have diffeomorphism invariance, and are built on usual manifolds. Higher spin theories on the other hand have a much bigger gauge invariance \cite{VasilievLatestReview}, big enough perhaps to make the notion of singularities and horizons gauge-dependent \cite{CK}. So it seems unlikely that our theories have a simple relation to higher spin theories. Nonetheless, it will be interesting to see if these theories have any relationship with Chern-Simons theories or Vasiliev theories, at least in specific cases and/or low dimensions.
}, we will instead look for another perspective on the Levi-Civita condition which might be more suitable for generalization.

That perspective is provided by the Palatini approach to general relativity where the connection and metric are treated as independent fields. In this formalism, we will view\footnote{This viewpoint is hardly original, Albert E himself tried this approach during his many attempts to get to his equations, well before Palatini.} the relationship between the two fields as a consequence of the choice of action: the specific choice of the Einstein-Hilbert action gives rise to the standard Levi-Civita condition as the connection equation of motion (when torsion is zero). This provides us with a natural avenue for generalizing (\ref{Levi-Civita}), by generalizing the action to include higher tensors. Note that the Einstein-Hilbert action had the feature that the connection was determined algebraically: Levi-Civita condition contains no derivatives on $\Gamma^a_{bc}$. We are not guaranteed that this will remain true if we generalize the action, and indeed it doesn't hold even in pure gravity if one includes higher derivative terms \cite{Sotiriou, Shahin, Janssen}. For a general action the connection is determined in terms of its (differential) equation of motion. We will see later that there exist classes of actions (which have Riemann appearing linearly) where the connection is determined algebraically in terms of the metric and higher tensors. For simplicity, the explicit examples in this paper are all taken to be of this form. The Einstein-Hilbert action is a special case in this class where the dependence is only on the metric. 

In any event, for now we will drop the algebraic determination of the connection as a strict requirement, and instead look at the equations of motion (algebraic or differential) from a Palatini formulation as the definition of our connection.

So our task then is to write down diff-invariant actions\footnote{We will mostly be concerned with vacuum situations in this paper (to the extent that the higher rank fields can be thought of us part of geometry), but it is straightforward to include matter by adding matter pieces covariantly coupled to the action. The distinction between matter and geometry is most natural in the cases where the connection is determined algebraically by its EOM, so that one can integrate it out from the action, and {\em then} couple the system to matter. Otherwise, the connection EOM will contain matter pieces. As an aside, we observe that in the case of both scalar and vector fields, the structure of the Lagrangian is such that the affine connection doesn't show up \cite{Sotiriou}. Also, it is perhaps worth being judicious about distinguishing GR experimental tests that are sensitive to the minimal coupling of the field as opposed to merely its geodesic (aka particle/WKB) limit. As theoretical pastime, one might also wish to study a fully geometric theory of matter, where one treats all fields on an equal footing. } containing the metric, higher tensors and the connection. Variations with respect to each of these fields will give us equations of motion, which together will define a geometrical theory of ``higher" gravity.  To get an action, we need to construct an appropriate scalar Lagrangian density and a volume form out of these fields, and integrate the former using the latter over the manifold. Furthermore, this Lagrangian density needs to contain derivatives, because we want to get differential (as opposed to algebraic) equations of motion for the tensors. In a manifold with an affine connection, we automatically have a good candidate tensor to help us do this: the Riemann tensor. When there are no tensors besides the metric, the natural action to write down at lowest order in derivatives is the Einstein-Palatini action
\begin{eqnarray}
S_{EH}[g, \Gamma]=\int d^4x \sqrt{g} \ g^{ab} R_{ab}(\Gamma).
\end{eqnarray}
Note that the definition of both the standard Riemann tensor and the Ricci tensor do {\em not} require the metric. But construction of the Ricci scalar does require the metric for contracting the indices. In principle we could consider higher powers and covariant derivatives of Riemann (which again would only depend on the connection) and construct scalars by appropriate contractions with the metric, and we would have a well-defined starting point for the variational principle. Typically this will then lead to higher derivative  equations of motions.

To include higher tensors (we will call them $\phi_{a_1a_2...a_n}$), we need two ingredients then. Firstly we need to be able to construct volume forms, and the second is we need to be able to write down scalar Lagrangians by contracting indices appropriately. We turn to the volume form first. Surely, we already have a well-defined volume form in 
\begin{eqnarray}\label{mvol}
\sqrt{g}\  dx^1 \wedge ...\wedge dx^d. 
\end{eqnarray} 
But we can do something more drastic. We can build our volume forms in ways that depend on the higher tensors as well. One of our goals in this paper is to see how much of the ``geometrodynamic" perspective on gravity is reliant on the metric. Put another way, general relativity tells us that rank two tensors can have an interpretation as geometry: we would like to see if other fields  (``matter") can in fact be geometrized in some sense as well. So we would like an approach that can stand on its own even when there is no metric. 

The defining property of the volume form is that it is a nowhere vanishing top form on the manifold, and that under diffeomorphisms, it transforms trivially (ie., upto a sign on an orientable manifold). Under a diffeomoprhism $x \rightarrow x'$, the form $dx^1 \wedge dx^2\wedge dx^3 \wedge dx^4$ transforms by a factor of ${\rm det} (\Lambda)$ where $\Lambda^{a'}_{b}=\partial x^{a'}/\partial x^b$. The factor of $\sqrt{g}$ is there to compensate for this in (\ref{mvol}). This is accomplished because the determinant of the metric
\begin{eqnarray}\label{mdet}
g= \frac{1}{4!}\epsilon_{a_1b_1c_1d_1}\epsilon_{a_2b_2c_2d_2}\ (g_{a_1a_2}g_{b_1b_2}g_{c_1c_2}g_{d_1d_2}),
\end{eqnarray}
transforms precisely by a compensating (square of) ${\rm det} (\Lambda)$. Note that the epsilons here stand for the epsilon symbol: they are {\em not} the Levi-Civita tensor and do {\em not} contain metric. The $4!$ here is a useful convention that corresponds to the dimensionality of spacetime which we have taken (for concreteness) to be 4. In $d$-dimensions, the determinant will be defined with a factor $\frac{1}{d!}$, there will be two (note that this two is the rank of the metric) epsilon tensors with $d$ legs each and there will be $d$ copies of the metric tensor. 

Now it is {\em trivial} to generalize this construction of such a volume form to a rank-$r$ tensor. In $d$-dimensions, the generalization of the determininant will be defined with a factor of $\frac{1}{d!}$, there will be $r$ epsilon tensors with $d$ legs each and there will be $d$ copies of the rank-$r$ symmetric tensor $\phi^{a_1...a_r}$. I.e., if one defines
\begin{eqnarray}\label{phidet}
\phi_r\equiv \frac{1}{d!}\epsilon_{a^1_1a^2_1...a^d_1}\dots \epsilon_{a^1_ra^2_r...a^d_r}\  \phi_{a^1_1a^1_2...a^1_r}
\dots \phi_{a^d_1a^d_2...a^d_r}
\end{eqnarray}
then it transforms with a factor of $({\rm det} \Lambda)^r$ under $GL(d,R)$ so that 
\begin{equation}
(\phi_r)^{1/r}\ dx^1\wedge...\wedge dx^d
\end{equation}
is a perfectly acceptable volume form. We adopt the convention that the rank-$r$  hyperdeterminant in (\ref{phidet}) is denoted by $\phi_r$, with $\phi_2 \equiv g$. We can generalize this further and construct more general volume forms based on various symmetric tensors on the manifold by combining these:
\begin{eqnarray}
dV=\prod_i(\phi_{r_i})^{q_{r_i}}\ dx^1\wedge...\wedge dx^d \label{genvolform}
\end{eqnarray}
This will be a well-defined volume form iff $\sum_i r_i q_{r_i}=1$.  
The existance of such general volume forms that allow for much more general {\em geometric} couplings between various symmetric tensors, is a simple yet key observation of this paper. Of course, if we have multiple tensors of the same rank, a slight generalization of the above volume form is possible, but we will not present it explicitly to avoid too much clutter. 

There is one caveat: The object (\ref{phidet}) above only exists for even $r$. For odd $r$, the quantity identially vanishes\footnote{This is easy to see using arguments of symmetry and anti-symmetry. One can also check it trivially for the special case when only the diagonal entries $\phi_{aa\dots a}$ (no summation) are no-zero.}. But a simple way to adapt our previous definition to the odd rank case is to first construct a fully symmetric even rank object out of it. For example for $\phi_{abc}$, we can construct $\Phi_{abcdef}=\phi_{(abc}\phi_{def)}$ which is even rank. This symmetrization approach works for any rank. Now we can define $(\Phi_{2r})^{1/2r}\ dx^1\wedge...\wedge dx^d$ as our volume form. We will mostly be dealing with even rank tensors in this paper.

The construction of the so-called Cayley hyperdeterminant for the specific case of rank 3 and dimension 2 has been discussed in \cite{Duff} in the context of black holes and qubits. We have stumbled on this object by the apporach above of looking at the transformation properties of the volume form, but we can use the explicit expression in \cite{Duff} as a sanity check: indeed, upto a numerical factor normalization the expression (4.2) in \cite{Duff} is the same as our construction for rank three and dimension two. The even rank case is more well-known\footnote{This object seems to have got some (but not too much) attention in mathematics, one reference that is quoted in a few places on the web is \cite{GKZ}. But \cite{GKZ} seem  to be dealing mostly (exclusively?) with what is called the ``geometric" hyperdeterminant, what we are dealing with here is a generalization of the ``combinatorial" hyperdeterminant, eg.  \cite{lim}.}, and we have checked that the object does not vanish.

That the volume forms we have constructed are ``good" objects to construct diff-invariant theories will become even more evident, when we vary our action to construct equations of motion. We will see that the resulting equations of motion can all be combined into manifestly covariant forms, which is a consequence of the diff-invariance of our action.

Armed with these volume forms, now we go on to construct scalar Lagrangian densities. 
But what are the independent tensors available to us for doing these contractions? For concreteness lets consider the case where the metric and a rank-3 tensor are the fields in the theory. One can of course invert the metric,
\begin{eqnarray}\label{min}
g^{ac}g_{cb}=\delta^a_b
\end{eqnarray}
and get the inverse metric as a possible tensor, because the right hand side is a well-defined invariant object under the coordinate transformation matrix $GL(d, R)$. But if we are allowed to use only the inverse metric as an object that can raise indices, it is easy to see that one can never construct scalars in a theory where there are only higher rank (covariant) tensors and {\em no} metric \footnote{We restrict ourselves to covariant symmetric tensors because keeping in touch with Riemann's original philsophy, we want to interpret them in terms of a notion of distance, eg. $ds^3 \sim \phi_{abc} dx^a dx^b dx^c$. We will soon see that one can construct contravariant higher rank tensors from covariant ones without resorting to metric, so this restriction is not really a restriction.}. 

A natural thing to try generalizing from (\ref{min}) is something like
\begin{eqnarray}
\phi^{abe}\phi_{ecd}\sim \delta^{(ab)}_{(cd)},
\end{eqnarray}
which would again define a well-defined tensor $\phi^{abe}$ (if it exists) because again the RHS is a $GL(d, R)$ invaraint. However the trouble is it doesn't exist in general:  the system is overconstrained as one can easily see by counting the number of equations, which is $\frac{d^2(d+1)^2}{4}$, and the number of independent degrees of freedom in $\phi$, which is $\frac{d(d+1)(d+2)}{6}$. This leads to trivial solutions in $d \ge 2$. The situation only gets worse if we go beyond rank 3 and consider even higher rank tensors. The basic problem is that in the case of the metric, since the product of two $d \times d$ matrices was a $d \times d$ matrix, multiplication had closure. Here we don't. So what are we to do?

A hint is provided by variations of the volume form. From usual differential geometry, we are familar with the relation
\begin{eqnarray}
g^{-1} \delta g= \ g^{a b}\ \delta g_{ab}.
\end{eqnarray}
The left hand  side is a tensor because the right hand side is. Now, explicitly evaluating the LHS and noting that $\delta g_{ab}$ are arbitrary variations, in 4 dimensions we get the relation
\begin{eqnarray}
g^{ab}=\frac{g^{-1}}{3!}\epsilon_{aa_1b_1c_1}\epsilon_{ba_2b_2c_2}g_{a_1a_2}g_{b_1b_2}g_{c_1c_2}.
\end{eqnarray}
This gives us an alternate definition of $g^{ab}$ indepedent of (\ref{min}), which we can try to generalize to our higher determinants. Explicitly evaluating $\phi_r^{-1}\delta \phi_r$ we get
\begin{eqnarray}
\phi_r^{-1}\delta \phi_r&=&\frac{\phi_r^{-1}}{(d-1)!}\epsilon_{a_1 a^2_1\dots a^d_1}\dots\epsilon_{a_r a^2_r\dots a^d_2}\times \nonumber \\ 
&& \hspace{1in}\times \phi_{a^2_1\dots a^2_r}
\dots \phi_{a^d_1\dots a^d_r}\ \delta \phi_{a_1 \dots a_r}\nonumber \\
&\equiv & \  \phi^{a_1 \dots a_r}\delta \phi_{a_1 \dots a_r}  \label{invdef}
\end{eqnarray}
The object $ \phi^{a_1 \dots a_r}$ is perfectly well-defined, the question is if it is a tensor. But from our construction of the volume form we already know that $\varepsilon^{\phi}_{a_1\dots a_d}\equiv \phi_r^{1/r} \epsilon_{a_1\dots a_d}$ and its inverse $\varepsilon^{a_1\dots a_d}_{\phi} = \phi_r^{-1/r} \epsilon_{a_1\dots a_d}$ are tensors\footnote{We have put $\phi$-sub/super-scripts on $\varepsilon$ to emphasize that they correspond to the volume forms constructed from the higher tensor and not the metric.}, so 
\begin{eqnarray}
\phi^{a_1 \dots a_r}\equiv \frac{1}{(d-1)!}\varepsilon^{a_1 a^2_1\dots a^d_1}\dots\varepsilon^{a_r a^2_r\dots a^d_2}
\phi_{a^2_1\dots a^2_r}
\dots \phi_{a^d_1\dots a^d_r} \nonumber \\
\end{eqnarray} 
is a tensor, which is exactly what definition (\ref{invdef}) is. So we have managed to obtain an upper index object purely from the higher tensor without using the metric, which can be used for constructing scalars. 

So now a general theory that we can write down using these ingredients will consist of a Lagrangian density that is a scalar constructed from 
\begin{eqnarray}
\phi_{a_1\dots a_{r_i}}, \ \  \phi^{a_1\dots a_{r_k}}, \ \ \nabla_a, \ \ R^a_{bcd}, \ \ R_{bd}
\end{eqnarray}
where there can be tensors (including the metric) of various ranks and their ``inverses"\footnote{The objects defined in (\ref{invdef}) can be thought of as inverses in that one can check that $ \phi^{a_1\dots a_{r_{k-1}}a}\phi_{a_1\dots a_{r_{k-1}}b}=\delta^{a}_{b}$. Note however that only for the rank 2 case is it a true inverse.}, the covariant derivative $\nabla_a$ is taken with the connection $\Gamma^a_{bc}$ which is thought of as an independent field, and the Riemann and Ricci are obtained as usual from the connection (and do not require the metric for their definition). 

Integrating Lagrangian densities of this kind over volume forms of the kind (\ref{genvolform}) gives us theories that generalize the standard Palatini construction of Einstein's gravity. Schematically, we can write such a theory as
\begin{eqnarray}
S=\int dV\ {\mathcal L}(\phi_{a_1\dots a_{r_i}}, \ \phi^{a_1\dots a_{r_k}}, \ \nabla_a, \  R^a_{bcd}, \ R_{bd})
\end{eqnarray}
where all the indices in $\mathcal L$ are understood to be contracted so that it is a scalar. Note that the connection is treated as an independent field as well, and is to be varied with\footnote{This is legitimate because even though the connection is not a tensor, its variation is, and therefore equations of motion arising from connection variation are tensor equations. This is a familiar fact: even though we vary the Maxwell action with respect to a gauge-dependent quantity (the gauge field $A_\mu$), we end up getting Maxwell equations which are gauge-invariant.}.

Of course there is always the possibility of coupling vaious terms of this kind together, as well as coupling these terms to covariantized matter: some interesting directions of this kind will be pursued in a forthcoming publication. 

Next, we turn to a few simple explicit example theories of this kind.

\section{Examples}

Lets first introduce a 0-tensor (scalar) together with the metric. The simplest action is of the form
\begin{equation}
S_=\int d^4x e^{\phi}\sqrt{g} \ g^{ab} R_{ab}
\end{equation}
The fact that the volume form is nowhere-vanishing immediately leads to an $e^{\phi}$ structure to its scalar part, leading to a dilaton-like structure. The dilaton equation of motion leads to $g^{ab} R_{ab}=0$, and the metric equation of motion leads to the usual Einstein equation, both together leading to $R_{ab}=0$. But note that the connection is determined by its own equation of motion and is no longer Levi-Civita. It is determined by 
\begin{eqnarray}
\nabla_ag^{bc}=\frac{2}{d-2} (\nabla_a\phi) \ g^{bc}  \label{nonLCdil}
\end{eqnarray}
The connection can be algebraically computed, and using the usual cycling of the indices and adding an subtracting, one can find its explicit form, but we will not present it here. As it stands the above theory does not have equations of motion for the scalar, if we do not couple the action above to something else (say a standard scalar kinetic term). This is a generic feature when a field appears only in the volume form and not in the Lagrangian density. Note however that once we go from a Palatini formulation to a metric-like formulation by explicitly solving for the connection and plugging back in, we do have dilaton derivative terms in our action because our connection (\ref{nonLCdil}) is scalar-dependent\footnote{It is conceivable that one way to look at our theories is precisely after doing this Palatini to metric-like translation. Note that once this is done, the coupling to non-geometric matter will proceed as usual: via covariantizing with respect to the connection. Of course, this is most natural when the connection is determined algebraically, as in the examples we present here. It will also be interesting to consider theories where all fields are geometrical, as in, they contribute to the volume form. That will be a fully geometrized theory of matter! If one wants to include spinors as well, the natural context to consider such theories would be in the context of (generalized) Riemannian supermanifolds \cite{Bryce}. It is clear that in our set-up, because of the presence of a Lorentzian metric, it is striaghtforward to couple spinors via the intreoduction of a local frame basis \cite{fu1}. But Cartan's structure equations will have to be re-considered.}!

Other possible theories include new types of bi-metric theories (see \cite{Clifton} for a review of bi-metric theories in cosmology):
$\int d^4x(g)^{1/p} (\tilde g)^{1/q} \ g^{ab} R_{ab}$, $\int d^4x(g)^{1/p} (\tilde g)^{1/q} \ {\tilde g}^{ab}{\tilde g}^{cd}g_{ab} R_{cd}$ etc are examples. Note that here $1/p+1/q=1/2$. 

As long as the Riemann tensor appears without covariant derivatives, the connection is determined algebraically. And it is determined linearly in terms of field derivatives when the action is linear in the Riemann (or Ricci) tensors. Note that we have specifically chosen such actions here for illustration, but there is nothing that prevents one from considering more complicated actions with mode derivatives and powers of the curvatures.

We will present one more theory (richer than the ones we considered so far) before concluding. This one contains a metric and a higher rank field. Note first that the higher fields have to have non-zero vev in order for the volume form to be well-defined, but a non-zero vev will typically break Lorentz invariance. This is unavoidable if we are wroking with odd rank fields, but for even rank fields we can in principle have vevs for the fields which are constructed from $\eta_{ab}$. So we will for the moment limit ourselves to even rank fields. An interesting theory is
\begin{eqnarray}
S=\int d^dx (g)^{1/q}(\phi_4)^{1/p} \phi^{abcd}g_{ab}R_{cd} \label{phi4theory}
\end{eqnarray}
with $2/p+1/q=1/2$. The equations of motion are
\begin{eqnarray}
g_{(ab}R_{cd)}-\frac{1}{p}\phi_{abcd}(\phi^{pqrs}g_{pq}R_{rs})=0, \\
\phi^{abcd}R_{cd}+\frac{1}{q}g^{ab}(\phi^{pqrs}g_{pq}R_{rs})=0, \\
\nabla_a \beta^{bc}+\beta^{bc}\Big(\frac{1}{q}g^{mn}\nabla_ag_{mn}+\frac{1}{p}\phi^{pqrs}\nabla_a\phi_{pqrs}\Big)=0.
\end{eqnarray}
where $\beta^{ab}\equiv \phi^{abcd}g_{cd}$, it is a symmetric rank-2 matrix whose inverse $\beta_{ab}$ is also useful. The last equation is the connection equation of motion, and it algebraically determines the connection:
\begin{eqnarray}
\Gamma^q_{a d}=\frac{1}{2}\beta^{qp}\Big(\beta_{pd,a}+\beta_{pa,d}-\beta_{a d,p} + \hspace{0.3in}\nonumber\\
\hspace{0.5in} +\alpha_{a}\beta_{pd}+\alpha_d \beta_{a d}- \alpha_p \beta_{d a}\Big)
\end{eqnarray}
Here $\alpha_a$ is defined by 
\begin{eqnarray}
\alpha_a=\frac{\beta_{bc}\partial_a \beta^{bc}}{d-2}-\frac{2}{d-2} \Big(\frac{1}{q}g_{mn}\partial_ag^{mn}+\frac{1}{p}\phi_{pqrs}\partial_a\phi^{pqrs}\Big). \nonumber \\
\end{eqnarray}
In many ways, this theory is a direct generalization of the Einstein-Hilbert action to include both rank 2 and rank 4 tensors. It will be interesting to investigate whether this connection here can be obtained from a curve length minimization\footnote{It should be emphasized that affine parellelism and distance minimization give rise to two different notions of geodesics, and Levi-Civita happens to be a case where the two coincide. But it is not clear to me that it is the only possibility. I thank Pallab Basu for an interesting discussion on this.} like the Levi-Civita connection arises from proper-time minimization\footnote{A potential generalization, for example, could be $\int_C \ (g_{ab}dx^adx^b)^{1/q}(\phi_{pqrs}dx^pdx^qdx^rdx^s)^{1/p}$.}. We will discuss some physical aspects and solutions of this theory in a follow-up paper.

Note that in principle we can work with much more general theories than what we have presented here, involving higher powers and higher derivatives. If one works with purely higher rank tensors without metric, one will typically have to deal with higher derivative equations of motion. 


\section{Discussion}

One question is if some of these theories are of phenomenological viability. It is tempting to speculate that they might be relevant for dark matter/energy. To have an idea about this, we will first have to construct (cosmological, black-hole-like,...) solutions in specific theories, which seems certainly doable. Finding the Newtonian/quadratic limit of our actions will also be useful for identifying theories which might have phenomenologial viability \cite{fu1}. Examples of gravity modification for dark sector phenomenology can be found in \cite{GravMod}.

Our theories have a Lorentzian metric in them, and one can always write them in a manifestly locally Lorentz invariant way by introducing local frames. We can also write locally Lorentz invariant matter couplings via frames, coupling fields of various spin. These actions will look manifestly local Lorentz invariant, and this is the way one imagines coupling standard quantum field theory to a curved geometry. 

A generic solution of these theories will break Lorentz invariance spontaneously\footnote{A bit more severely than in pure GR because in GR we always have the option of turning off the fields other than the metric, and in such solutions {\em local} Lorentz invariance is not broken.}. But the vacuum can be Lorentz invariant: in theory (\ref{phi4theory})  consider $g_{ab}=\eta_{ab}$ with  $\phi_{abcd}= \alpha (\eta_{ab}\eta_{cd}+ \eta_{ac}\eta_{bd} + \eta_{ad}\eta_{bc})$. There are three terms in this last expression and together they make the expression fully symmetric and $\alpha$ is the vev of the field. It is easy to check that this solves the equation of motion: we have therefore constructed a Lorentz-invariant vacuum, for each value of $\alpha$. Note that setting $\phi_{abcd}=0$ is not  acceptable because $\phi$ shows up in the volume form. Note also that this type of non-trivial vacuum cannot be found with odd-rank fields, because giving them a non-trivial vev will break Lorentz. 

However there is a subtelty here that is not there in standard general relativity. Even though we can make the theory manifestly locally Lorentz invariant by erecting local frames, the causal structure is not disctated by lightcones. The reason is easy to understand: the Levi-Civita condition is precisely the condition that guarantees that the proper-time element remains invariant along a geodesic. In our case the connection depends on derivatives of the higher field(s) as well, and therefore the notion of causality is no longer only based on the metric and lightcone. A somewhat similar situation happens in Born-Infeld theory where there is tension between two notions of causality \cite{Gibbons}.

This seems like a new type of example for what usually falls under the rubric of Lorentz violation, and might be worth exploring: typically, breaking of Lorentz invariance is due to the addition of explicit terms in the Lagrangian that break it, or spontaneously in the vacuum. Here, on the other hand a potential superluminality shows up because the parallel transport of the frames is not metric compatible. Note however that this does {\em not} necessarily mean that the theory is inconsistent, the notion of what causality means is modified in these theories\footnote{This doe not mean however that it is phenomenologically viable. If one plans to use the higher rank fields as some form of dark matter, then either one will have to make sure that the solutions one considers do not allow superluminality, or (to play devil's advocate) will have to come up with observational evidence that in regions where there is dark matter gradients or boundaries, one has the required (hopefully mild) form of superluminality. To push this last speculation a bit further, one might ask whether the observed bounds on Lorentz invariance and absence of superluminality are comparable in magnitude to the bounds on the dimensionless gradients of dark matter (say, in the solar system).}. For example, the usual argument that there exists Lorentz frames in which superluminal propagation looks backward in time is strictly speaking true only when the world is globally Lorentz invariant, so that there is sense in comparing the propagated event with the original event using local Lorentz transformations at the {\em same} point. Metric compatibility is what lets us extend this over different points on the manifold in standard GR. And here, we don't have  metric compatibility. Clearly, the issue of causality requires further study.



Another direction is to consider these theories in various dimensions. String theory is the interesting object to study in two dimensions using the volume form constructed from a worldsheet 2-tensor, the metric. In other dimensions and for other objects, it might be interesting to consider other rank tensors or combinations of them.


Note also that the question of what kind of theories of this kind (if any) are ghost-free is very interesting. There are various subtle issues related to higher spin fields which evidently need to be clarified \cite{Bekaert}. Part of the purpose of the present note was reconnaissance of theory-space. At the moment, we suspect that these theories have similarities to (higher generalizations of) massive gravity in the massless limit \cite{Hint} (as obtained via a Stuckelberg trick), but we will postpone discussions in this direction to a different paper \cite{fu1}.

What is the point of all this? For one, it is interesting that a whole class of theories, closely related to and generalizing Einstein-Hilbert gravity, exists. The interest in these theories is at first blush theoretical, but it could also have some phenomenological viability in making IR modifications to gravity. The ease with which all the constructions fell into place here is surprising, and it is evident that there is a whole plethora of questions that need answering. \\

I thank Aiyalam Parameswaran Balachandran, Pallab ``No-Longer-Kumar" Basu and Mudumbai Seshachalu Narasimhan for discussions and/or encouragement and Payingattery Natarajan Bala Subramanian for a related collaboration.

\end{document}